\documentclass[a4paper,twoside]{amsart}

\usepackage[english]{babel}
\usepackage{amsmath}
\usepackage{amsfonts}
\usepackage{amsthm}
\usepackage{amssymb}
\usepackage{mathrsfs}			
\usepackage[shortlabels]{enumitem}
\usepackage{tikz}						
\usepackage[utf8]{inputenc}		
\usepackage{hyperref}

\usepackage{todonotes}

\pdfoutput=1

\allowdisplaybreaks		

\setlist[2]{labelsep = .5\parindent}
\setlist[3]{labelsep = .5\parindent}

\newcommand{\IN}{\mathbb{N}}

\newcommand{\fk}{\mathfrak{k}}

\newcommand{\cA}{\mathcal{A}}
\newcommand{\cH}{\mathcal{H}}
\newcommand{\cG}{\mathcal{G}}

\newcommand{\set}[2]{\{#1 \colon #2 \}}

\newcommand{\Id}{\text{\normalfont Id}}

\newcommand{\dom}{\text{\normalfont Dom}}

\usepackage[all]{xy}

\usepackage{color}

\newcommand{\A}{A} 

\theoremstyle{plain}
\newtheorem{thrm}{Theorem}
\newtheorem{lem}[thrm]{Lemma}
\newtheorem{prop}[thrm]{Proposition}

\theoremstyle{definition}

\newtheorem{rem}[thrm]{Remark}

\title[Reduction of quantum systems and the local Gauss law]{Reduction of quantum systems \\and the local Gauss law}
\author{Ruben Stienstra}
\author{Walter D. van Suijlekom}
\address{Institute for Mathematics, Astrophysics and Particle Physics, Radboud University Nijmegen, Heyendaalseweg 135, 6525 AJ Nijmegen, The Netherlands}
\email{r.stienstra@math.ru.nl, waltervs@math.ru.nl}

\date{\today}

\begin{document}


\begin{abstract}
We give an operator-algebraic interpretation of the notion of an ideal generated by the unbounded operators associated to the elements of the Lie algebra of a Lie group that implements the symmetries of a quantum system. We use this interpretation to establish a link between Rieffel induction and the implementation of a local Gauss law in lattice gauge theories similar to the method discussed by Kijowski and Rudolph in \cite{KR02,kijowski04}.
\end{abstract}

\maketitle

\section{Introduction}

\noindent
There are well-developed theories of reduction of both classical and quantum mechanical systems that possess symmetries.
The study of reduction of classical systems was initiated by Dirac in \cite{dirac50} with his theory of first and second order constraints, and later put into the language of symplectic manifolds by Arnold and Smale.
The reduction of a symplectic manifold with respect to an equivariant moment map was described by Marsden and Weinstein in their paper \cite{marsden74}.
For a more detailed account of the history of symplectic reduction, we refer to \cite{marsden01} and references therein.
A procedure known as Rieffel induction, developed in \cite{rieffel74}, appears to be a good candidate for a quantum version of Marsden--Weinstein reduction \cite{landsman95} (cf. \cite[IV.2]{landsman98}).

The primary aim of this paper is to compare two different ways to reduce the quantum mechanical observable algebra.
The first one is the method of Rieffel induction mentioned above.
The second one was outlined by Kijowski and Rudolph in \cite{KR02,kijowski04} in the context of a quantum lattice gauge theory, in which they explicitly implement a constraint, the local Gauss law, by ensuring that the operators associated to the generators of the gauge group vanish in the observable algebra of the reduced system.
The corresponding operators on the unreduced Hilbert space are unbounded, however, which requires them to appeal to the theory of C$^\ast$-algebras generated by unbounded operators as developed by Woronowicz in \cite{woronowicz95}, something that is not necessary for Rieffel induction.
Nevertheless, both procedures yield the same reduced observable algebra.
In this paper, we modify the latter method so that it is formulated entirely in terms of bounded operators, and show that it agrees with the final step in the process of Rieffel induction.

The paper is organised as follows.
In Section 2, we briefly recall the process of Rieffel induction.
In Section 3, we formulate and prove the main theorem that establishes the link.
In Section 4, we discuss some examples, including the lattice gauge theory mentioned above.

\subsection*{Acknowledgements}	 We would like to thank Francesca Arici, Klaas Landsman and Gerd Rudolph for useful comments and helpful discussions. This research was partially supported by NWO under the VIDI-grant \mbox{016.133.326}. 

\section{Reduction of quantum systems using Rieffel induction}
\label{sec:Reduction_of_quantum_systems}

\noindent 
The kinematical data of a quantum system consists of a Hilbert space $\cH$ and a faithful representation $\pi$ of a C$^\ast$-algebra $\A$ on $\cH$.
A continuous symmetry of a quantum system typically (but, in accordance with Wigner's celebrated theorem, not exclusively,) corresponds to a continuous unitary representation $\rho \colon K \rightarrow U(\cH)$ of some Lie group $K$ on $\cH$.
We are interested in studying the reduction of the kinematical data with respect to such symmetries in the case in which $K$ is compact.
A systematic way to obtain this reduction, known as Rieffel induction, was proposed by Landsman in \cite{landsman95} using an induction procedure for representations of C$^\ast$-algebras developed by Rieffel in \cite{rieffel74}.

Let us first briefly recall the process of Rieffel induction.
Starting from the above representation of the group $K$, one endows $\cH$ with the structure of a right Hilbert $C^\ast(K)$-module, where $C^\ast(K)$ denotes the group C$^\ast$-algebra of $K$.
Subsequently, one takes the quotient of $\cH$ with respect to the null space of a bilinear form on $\cH$, which yields a space naturally isomorphic to $\cH^K$, the subspace of $\cH$ of $K$-invariant elements.
Thus we obtain the Hilbert space of the reduced system.

At the level of the observable algebra, one first considers the algebra $\A^K$ of elements of $\A$ that are equivariant with respect to the given unitary representation.
The space $\cH^K$ is invariant under these observables, yielding a representation $\pi$ of the C$^\ast$-algebra $\A^K$ on $\cH^K$.
The image of this representation is isomorphic to $\A^K/\ker(\pi)$, and hence one obtains a faithful representation of $\A^K/\ker(\pi)$ on $\cH^K$, which forms the remaining part of the kinematical data of the reduced system.

Motivated by the theory of strict quantization of observable algebras as described extensively in \cite[Part II]{landsman98}, we are interested in the case where $\A = B_0(\cH)$, the space of compact operators, and its representation on $\cH$ is the obvious one.
It can then be shown that $\A^K/\ker(\pi)$ is isomorphic to $B_0(\cH^K)$, and that the representation of this algebra on the reduced Hilbert space 
$\cH^K$ is again the obvious one.

\section{Associating algebras to infinitesimal generators}
\label{sec:algebra_generated_by_unbounded_elements}

\noindent
The main purpose of this section is to discuss a possible interpretation of an observation made by Kijowski and Rudolph in \cite[Section 3]{kijowski04} in the case of a quantum lattice gauge theory, namely that the kernel of the representation $\pi \colon \A \rightarrow B(\cH)$, where as before $\A = B_0(\cH)$, is in some sense generated by the elements of the Lie algebra $\fk$ of the symmetry group $K$.
The representation of the group $K$ on $\cH$ can be used to associate differential operators to the elements of $\fk$, which are typically unbounded if $\cH$ is infinite-dimensional. If instead the representation space is finite-dimensional then the representation of the Lie algebra $\fk$ is bounded. 
Using this fact and other standard results from the representation theory of Lie groups, we will show how the differential operators associated to the elements of the Lie algebra generate $\ker(\pi)$.
In addition, we need the following preparatory lemma, which can be found in \cite[Exercise 4.2(c)]{murphy90}:

\begin{lem}
Let $\cH
$ be a Hilbert space, let $a$ be a compact operator on $\cH$, and suppose that $(b_j)_{j \in J}$ is a bounded net of bounded operators that converges strongly to $b \in B(\cH)$.
Then the net $(b_j a)_{j \in J}$ converges in norm to $ba$. 
If in addition the operator $b_j$ is hermitian for each $j \in J$, then the net $(ab_j)_{j \in J}$ converges in norm to $ab$.
\label{lem:strong+compact_implies_norm}
\end{lem}

\noindent
The following result shows how $\ker(\pi)$ can be generated by differential operators:

\begin{thrm}
Suppose $K$ is a compact, connected Lie group.
Let $S$ be a collection of finite-dimensional subrepresentations of the continuous representation $\rho \colon K \rightarrow U(\cH)$, and for each $\sigma \in S$, let $\cH_\sigma \subseteq \cH$ be the subspace on which $\sigma$ is represented.
Suppose that these representation spaces form an orthogonal decomposition of $\cH$, i.e.,
\begin{equation*}
\cH 
= \overline{\bigoplus_{\sigma \in S} \cH_\sigma}.
\end{equation*}
Then $\ker(\pi)$ is the closed, two-sided ideal generated by the set
\begin{equation}
\left\{ \int_{K} \rho(k) \sigma(X)^n \rho(k)^{-1} \: dk
\colon \sigma \in S, \: X \in \fk, \: n \geq 1 \right\}.
\label{eq:generators_of_the_kernel}
\end{equation}
\label{thrm:decomposition_and_algebra_generators}
\end{thrm}

\begin{rem}
In the set of generators above, $\sigma(X)$ is regarded as the compression of $\rho(X)$ to $\cH_\sigma$.
Moreover, we note that the integrals of vector-valued functions can be defined using Bochner integration.
\end{rem}

\begin{proof}[Proof of Theorem \ref{thrm:decomposition_and_algebra_generators}.]
Let $I$ be the ideal in $\A^K$ generated by the set in equation \eqref{eq:generators_of_the_kernel}.
We first show that $I \subseteq \ker(\pi)$.
Indeed, $\sigma(X)^n$ maps $\cH$ into $\cH_\sigma$ for each $\sigma \in S$, each $X \in \fk$ and each $n \geq 1$, hence so does $\int_{K} \rho(k) \sigma(X)^n \rho(k)^{-1} \: dk$, which implies that it is a finite rank operator.
In particular, it is compact.
Moreover, it follows from left invariance of the Haar measure that $\int_{K} \rho(k) \sigma(X)^n \rho(k)^{-1} \: dk$ is equivariant with respect to $\rho$, so it is an element of $\A^K$.
Finally, to show that it is an element of $\ker \pi$, let $p_\sigma \colon \cH \rightarrow \cH_\sigma$ be the orthogonal projection onto the representation space of $\sigma$. For each $v \in \cH^K$ we have $p_\sigma v \in \cH^K$ and $\sigma(X)v = 0$. Hence
\begin{align*}
 \int_{K} \rho(k) \sigma(X)^n \rho(k)^{-1}(v) \: dk 
= \int_{K} \rho(k) \sigma(X)^n (v) \: dk
= 0,
\end{align*}
and therefore $\int_{K} \rho(k) \sigma(X)^n \rho(k)^{-1} \: dk \in \ker(\pi)$.
Thus the generators of $I$ are contained in $\ker(\pi)$.
Since $\ker(\pi)$ is a closed, two sided ideal, it follows that $I \subseteq \ker(\pi)$.

We turn to the proof of the reverse inclusion.
Let $b \in \ker(\pi)$, let $p_{\cH^K}$ be the orthogonal projection of $\cH$ onto $\cH^K$.
It is easy to see that
\begin{equation*}
p_{\cH^K} = \int_{K} \rho(k) \: dk.
\end{equation*}
Since $b \in \ker(\pi)$, it follows that
\begin{equation*}
b
= b(\Id_{\cH} - p_{\cH^K})
= b \int_{K} ( \Id_{\cH} - \rho(k) )\: dk
= \sum_{\sigma \in S} b \int_{K} (p_\sigma - \sigma(k)) \: dk.
\end{equation*}
By the preceding lemma, the series on the right-hand side is norm-convergent, hence to show that $b \in I$, it suffices to show that
\begin{equation*}
b \int_{K} (p_\sigma - \sigma(k)) \: dk \in I,
\end{equation*}
for each $\sigma \in S$.
Since $I$ is closed under multiplication with elements of $\A^K$, we are done if we can show that
\begin{equation*}
\int_{K} (p_\sigma - \sigma(k)) \: dk \in I.
\end{equation*}
From bi-invariance of the Haar measure and Fubini's theorem, we infer that
\begin{equation*}
\int_{K} (p_\sigma - \sigma(k)) \: dg
= \int_{K} \int_{K} \rho(h) (p_\sigma - \sigma(k))\rho(h)^{-1} \: dh \: dk.
\end{equation*}
The norm topology and the strong topology coincide on the finite-dimensional algebra $B(\cH_\sigma)$, so the first integral on the right-hand side is a norm limit of Riemann sums, i.e. for each $\varepsilon > 0$, there exist $k_j \in K$ and $c_j \geq 0$ for $j = 1,\ldots,n$, such that
\begin{equation*}
\left\| \int_{K} \int_{K} \rho(h) (p_\sigma - \sigma(k))\rho(h)^{-1} \: dh \: dk
- \sum_{j = 1}^n c_j \int_{K} \rho(h) (p_\sigma - \sigma(k_j))\rho(h)^{-1} \: dh \right\| < \varepsilon.
\end{equation*}
Since $I$ is closed by definition, it suffices to show that
\begin{equation*}
\sum_{j = 1}^n c_j \int_{K} \rho(h) (p_\sigma - \sigma(k_j))\rho(h)^{-1} \: dh \in I.
\end{equation*}
We prove this by showing that
\begin{equation*}
\int_{K} \rho(h) (p_\sigma - \sigma(k))\rho(h)^{-1} \: dh \in I,
\tag{$\ast$}
\label{eq:in_the_ideal}
\end{equation*}
for each $k \in K$.
Now fix such a $k$.
Because $K$ is both compact and connected, the exponential map $\exp \colon \fk \rightarrow K$ is surjective, so there exists an $X \in \fk$ such that $k = \exp(X)$.
But $\sigma$ is a homomorphism of Lie groups, so
\begin{equation*}
\sigma(k)
= \sigma \circ \exp(X)
= \exp \circ \sigma(X)
= p_\sigma \sum_{j = 0}^\infty \frac{\sigma(X)^j}{j!}.
\end{equation*}
Thus
\begin{equation*}
p_\sigma - \sigma(k)
= -\sum_{j = 1}^\infty \frac{\sigma(X)^j}{j!}.
\end{equation*}
The map
\begin{equation*}
B(\cH_\sigma) \rightarrow B(\cH_\sigma), \quad 
a \mapsto \int_{K} \rho(h) a \rho(h)^{-1} \: dh,
\end{equation*}
is a linear operator on the finite-dimensional algebra $B(\cH_\sigma)$, hence it is norm-continuous, so
\begin{equation*}
\int_{K} \rho(h) (p_\sigma - \sigma(k))\rho(h)^{-1} \: dh
= -\sum_{j = 1}^\infty \frac{1}{j!} \int_{K} \rho(h) \sigma(X)^j \rho(h)^{-1} \: dh,
\end{equation*}
and the series on the right-hand side converges with respect to the norm on $B(\cH)$.
Each of the partial sums is an element of $I$, which implies that \eqref{eq:in_the_ideal} holds, as desired.
\end{proof}

\noindent 
In general, the set $S$ in the above theorem will not be unique.
Suppose that we are in the situation of the theorem, and that we are given a set $S$ satisfying the assumption.
If the Hilbert space $\cH$ is infinite-dimensional, there are infinitely many different sets like $S$ that satisfy the assumption.
Indeed, $S$ is an infinite set because $\cH$ is infinite-dimensional, so we can take any finite subset $F \subseteq S$ containing at least two representations, define the subrepresentation $\sigma_F := \bigoplus_{\sigma \in F} \sigma$, and the set $S^\prime = (S \backslash F) \cup \{\sigma_F\}$.
Then $S^\prime \neq S$, and it satisfies the assumption of the theorem.

The last argument can be formulated slightly more generally as follows:
Suppose that $S_1$ and $S_2$ are sets of orthogonal finite-dimensional subrepresentations, and that $S_1$ satisfies the assumption of the theorem.
If each element of $S_1$ is a subrepresentation of $S_2$, then $S_2$ also satisfies the assumption.
If $\cH$ is infinite-dimensional, then from any set $S_1$ one can always construct a different set $S_2$ with these properties.
Thus one can always make the set $S$ `arbitrarily coarse', which is another reason why we view Theorem \ref{thrm:decomposition_and_algebra_generators} as a possible way to make the idea of `the ideal generated by unbounded operators' rigorous.

The fact that a set $S$ like the one in Theorem \ref{thrm:decomposition_and_algebra_generators} always exists, is a consequence of the following result.
Recall that for any representation $\rho$ of a group $K$ on a space $V$, a vector $v$ is called {\em $K$-finite} if and only if the smallest subspace containing $v$ that is invariant under $\rho$, i.e., the span of $\set{\rho(k)v}{k \in K}$, is finite-dimensional.
We let $V^{\text{\normalfont fin}}$ denote the subspace of $K$-finite vectors of $V$.


\begin{prop}
Let $\rho$ be a continuous representation of a compact Lie group $K$ in a complete locally convex topological vector space $V$.
Then $V^{\text{\normalfont fin}}$ is dense in $V$.
\end{prop}

\noindent 
This result can be found in \cite{duistermaat00} as part of Corollary 4.6.3.
Using this result and Zorn's lemma, one can now readily show that there exists a set $S$ which satisfies the assumption of our theorem.
Needless to say, explicitly exhibiting such a set might be impossible.
However, as we shall see in the next section, there are situations in which there is a natural choice for $S$.

Before we end this section, we briefly recall some other notions from representation theory.
Let $\widehat{K}$ be the set of equivalence classes of irreducible representations of $K$, and let $[\delta] \in \widehat{K}$.
The {\em isotypical component of type $[\delta]$} is the set $V[\delta]$ of elements $v \in V^{\text{\normalfont fin}}$ such that the subrepresentation generated by $v$ is equivalent to the representation $\delta \oplus \dots \oplus \delta$ ($n$ copies) for some $n \in \IN$.

\section{Examples}
\label{sec:examples}

\subsection{Local Gauss law in quantum lattice gauge theories}
\label{ex:Gauss_law_lattice_gauge_theory}
We start with the motivating example for this paper, namely the local Gauss law discussed by Kijowski and Rudolph in \cite{kijowski04} in the context of quantum lattice gauge theories. Let $\Lambda = (\Lambda^0, \Lambda^1)$ be a finite, connected, oriented graph whose sets of vertices and edges are given by $\Lambda^0$ and $\Lambda^1$, respectively.
Moreover, let $s,t \colon \Lambda^1 \rightarrow \Lambda^0$ be the maps that assign to an edge its source and target, respectively.
Finally, let $G$ be a compact Lie group.
Let $\cG := G^{\Lambda^0}$, and let $\cA := G^{\Lambda^1}$ be the sets of functions from $\Lambda^0$ and $\Lambda^1$ to $G$, respectively.
In lattice gauge theory, $\cG$ is the gauge group, while $\cA$ is the space of connections.
We endow these sets of functions with Lie group structures simply by viewing them as direct products of $G$ with itself.
Then $\cA$ carries an action of $\cG$, which is given by
\begin{equation*}
(g_x)_{x \in \Lambda^0} \cdot (a_e)_{e \in \Lambda^1}
:= (g_{s(e)} a_e g_{t(e)}^{-1})_{e \in \Lambda^1}.
\end{equation*}
This action induces a continuous unitary representation $\rho$ of $\cG$ on $\cH := L^2(\cA)$ by $(\rho(g)(\psi))(a) := \psi(g^{-1} \cdot a)$, where $g \in \cG$, $\psi \in \cH$ and $a \in G^{\Lambda^1}$.

Baez already noted in \cite{baez96} that the action of $\cG$ restricts to the isotypical components of $\cH$ with respect to the left regular representation of $G^{\Lambda^1}$ on this space ---in fact, these isotypical components form the basis for spin networks as introduced in \cite{rovellismolin95}.
Indeed, the representation $\rho$ can be regarded as the composition of two group homomorphisms; first, we have a homomorphism
\begin{equation*}
\iota \colon \cG \rightarrow G^{\Lambda^1} \times G^{\Lambda^1} \simeq (G \times G)^{\Lambda^1}, \quad 
(g_x)_{x \in \Lambda^0} \mapsto (g_{s(e)}, g_{t(e)})_{e \in \Lambda^1},
\end{equation*}
which by connectedness of the graph is an injection if and only if $\Lambda$ has more than one vertex.
The second homomorphism is the product representation $L \times R \colon G^{\Lambda^1} \times G^{\Lambda^1} \rightarrow U(\cH)$ of the left and right regular representations $L$ and $R$, respectively.
It follows that each subspace of $\cH$ that is invariant under the representation $L \times R$, is also invariant under $\rho$.
The Peter--Weyl theorem asserts that
\begin{equation*}
\cH^{\text{\normalfont fin}} = \bigoplus_{[\delta] \in \widehat{G^{\Lambda^1}}} \cH[\delta],
\end{equation*}
and that the isotypical components $\cH[\delta]$ are irreducible subrepresentations of the representation $L \times R$ of dimension $\dim(\delta)^2$.
Here, $\cH^{\text{\normalfont fin}}$ denotes the set of $G^{\Lambda^1}$-finite vectors with respect to the left regular representation of $G^{\Lambda^1}$ on $L^2(\cA)$.
Thus we may take the set $S$ in Theorem \ref{thrm:decomposition_and_algebra_generators} to be the collection of subrepresentations obtained by restricting $\rho$ to $\cH[\delta]$ for each $\delta \in \widehat{G^{\Lambda^1}}$.

Since the elements of the Lie algebra of the gauge group $\cG$ generate the gauge group, Theorem \ref{thrm:decomposition_and_algebra_generators} provides a link between two different methods of reduction of the quantum observable algebra, the first being Rieffel induction, and the second being the implementation of a local Gauss law by taking the quotient with respect to an ideal generated by unbounded operators associated to Lie algebra elements, as mentioned by Kijowski and Rudolph in \cite{kijowski04}.

\subsection{Hamiltonian symmetries}
The second example that we discuss is really more of a class of examples, namely that of quantum systems with a given Hamiltonian that possesses a certain symmetry.

Let $\cH$ be a Hilbert space, let $H$ be a (possibly unbounded) self-adjoint operator on $\cH$, and suppose $\rho \colon K \rightarrow U(\cH)$ is a continuous unitary representation of a compact connected Lie group $K$ on $\cH$ with the property that $\rho(k)$ preserves $\dom(H)$ and $[\rho(k),H] = 0$ for each $k \in K$.
Moreover, let $\sigma_p(H)$ be the point spectrum of $H$, and for each $\lambda \in \sigma_p(H)$, let $\cH_\lambda$ be the eigenspace corresponding to $\lambda$.
Suppose that $\cH_\lambda$ is finite dimensional for each $\lambda \in \sigma_p(\lambda)$, and that $\cH = \overline{\bigoplus_{\lambda \in \sigma_p(H)} \cH_\lambda}$.
Then $\rho$ restricts to a representation $\rho_\lambda$ on $\cH_\lambda$ for each $\lambda \in \sigma_p(H)$, and we may set $S := \set{\rho_\lambda}{\lambda \in \sigma_p(H)}$.

A notable subclass of examples satisfying the above conditions is the class of quantum systems in which $\cH = L^2(Q)$, where $Q$ is a compact smooth Riemannian manifold that admits a Lie group of isometries, and $H = \Delta$ is the Laplacian on $Q$.
In particular, the lattice gauge theories in Section \ref{ex:Gauss_law_lattice_gauge_theory} can be studied in this way if one endows $G^{\Lambda^1}$ with a bi-invariant Riemannian metric.
It is a result from representation theory (cf. \cite[Theorem 3.3.5]{taylor86}) that $\cH[\delta]$ is a subspace of an eigenspace of $\Delta$ for each $\delta \in \widehat{G^{\Lambda^1}}$, so the decomposition obtained in the previous example is finer than the decomposition into eigenspaces of $\Delta$.


\end{document}